\documentclass[twocolumn,amsmath,amssymb]{revtex4}
\usepackage{graphicx}
\usepackage{dcolumn}
\usepackage{bm}
\begin{document}

\title{Tunability of the Fractional Quantum Hall States in Buckled Dirac Materials}
\author{Vadym M. Apalkov}
\affiliation{Department of Physics and Astronomy, Georgia State University,
Atlanta, Georgia 30303, USA}
\author{Tapash Chakraborty$^\ddag$}
\affiliation{Department of Physics and Astronomy,
University of Manitoba, Winnipeg, Canada R3T 2N2}

\date{\today}
\begin{abstract}
We report on the fractional quantum Hall states of germanene and silicene where one expects a strong 
spin-orbit interaction. This interaction causes an enhancement of the electron-electron interaction
strength in one of the Landau levels corresponding to the valence band of the system. This 
enhancement manifests itself as an increase of the fractional quantum Hall effect gaps compared to 
that in graphene and is due to the spin-orbit induced coupling of the Landau levels of the conduction
and valence bands, which modifies the corresponding wave functions and the interaction within a 
single level. Due to the buckled structure, a perpendicular electric field lifts the valley 
degeneracy and strongly modifies the interaction effects within a single Landau 
level: in one valley the perpendicular electric field enhances the interaction strength
in the conduction band Landau level, while in another valley, the electric field strongly suppresses 
the interaction effects.

\end{abstract}
\maketitle

The unique electronic properties of graphene \cite{graphene_book,abergeletal}, for 
example, that of the interacting Dirac fermions in an external magnetic field 
\cite{mono_FQHE,bi_FQHE,FQHE_chapter,interaction} have captivated our collective attention 
now for almost a decade. However, very recently, other emergent Dirac materials, such 
as silicene and germanene have rapidly gained considerable attention \cite{review} 
because of their even greater promises. These systems are similar in structure as 
that of graphene, and hence contain those remarkable properties of graphene, and 
then some \cite{first,yao,zhang}. They are monolayers of silicon and germanium with 
hexagonal lattice structures where the low energy charge carriers are also massless 
Dirac fermions \cite{silicene}. The interesting additional behavior lies in the {\it 
buckled} structure \cite{buckled} of these systems due to their larger ionic radius 
than that of carbon, whereby the two sublattices in these systems are displaced 
vertically. As a consequence, a large spin-orbit interaction (SOI) induced gap opens 
up at the Dirac points ($\Delta^{}_{\rm so} \approx 1.55 - 7.9$ meV for silicene 
\cite{yao} and $\Delta^{}_{\rm so} \approx 24 - 93$ meV for germanene \cite{yao}). 
This is in contrast to the tiny SO gap of about 25 $\mu$eV in graphene \cite{fabian}. 
The buckled structure of the lattice allows for the band gap to be tunable \cite{falko}. 
It has been suggested that with an applied perpendicular electric field the band gap 
can actually be controlled \cite{ezawa} as the size of the band gap increases 
linearly with the electric field strength. Quite naturally, this has generated a 
huge surge in interest in exploring the properties of Dirac fermions in these two 
systems, with an eye to their great potential for device applications. The fractional 
quantum Hall effect (FQHE) states of interacting Dirac fermions 
\cite{mono_FQHE,bi_FQHE,FQHE_chapter,interaction} 
are particularly intriguing in this context. The SOI is expected to significantly enhance 
the FQHE gap \cite{califano}. The large SO coupling in the present systems makes the 
FQHE states uniquely susceptible to an external control, and consequently a greater 
insight into the effect.

The FQHE in graphene has revealed some novel features specific to the relativistic systems. 
The electron-electron interactions are the strongest not in the $n=0$ Landau level (LL) as 
in conventional semiconductor systems, but in the $n=1$ LL, which results in the largest FQHE 
gaps in the $n=1$ LL \cite{mono_FQHE}. Here $n$ is the LL index. The wave functions in the $n=0$ 
LL in graphene are completely identical to the wave functions of the $n=0$ LL of a conventional 
(nonrelativistic) system. The buckled structure of silicene and germanene not only modify 
their energy spectrum from that of graphene but the strong SOI also lifts the spin degeneracy, 
while an external electric field lifts the valley degeneracy of the energy levels. In a magnetic 
field, the LL spectra and the corresponding wave functions can be also modified and controlled 
by the electric field. Here we study the effects of the electric field on 
the correlation properties of the Dirac fermions in the FQHE regime. The measure of the 
strength of the electron-electron interactions can be characterized by the magnitude of the 
corresponding FQHE gaps. For bilayer graphene, it was shown \cite{bi_FQHE,FQHE_chapter} 
that the bias voltage can strongly modify the property of the FQHE states and in some cases 
even increase the corresponding gap compared to that of the monolayer graphene. These effects 
are expected for monolayer silicene and germanene solely due to the SOI.

\begin{figure}
\begin{center}\includegraphics[width=7cm]{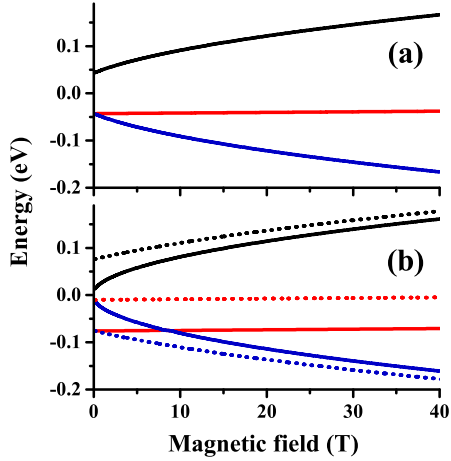}\end{center}
\vspace*{-1cm}
\caption{
Energies of three LLs corresponding to the Hamiltonian (\ref{HM}) as a function of the
perpendicular magnetic field. The perpendicular electric field is (a) $E^{}_z = 0$ 
and (b) $E^{}_z = 0.1$ V/\AA. For $E^{}_z = 0$, each level has twofold valley degeneracy. The 
degeneracy is lifted for a finite electric field [panel (b)], where the levels of the
$K$ valley are shown by solid lines, and the levels of $K^{\prime }$ valley are shown by dashed 
lines. The results shown here are for germanene.
}
\label{Fig_energy_E0_E01}
\end{figure}

The low-energy Hamiltonian of the silicine/germanene monolayer is given by \cite{ezawa} 
\begin{equation}
{\cal H}^{}_{\eta} = v^{}_F \left( p^{}_x \tau^{}_x - \eta p^{}_y \tau^{}_y \right) 
+ \eta \tau^{}_z h + L^{}_z E^{}_z \tau^{}_z,  
\label{H1}
\end{equation}
where 
\begin{equation}
h = - \lambda^{}_{\rm SO} \sigma^{}_z - a\hbar^{-1} \lambda^{}_R \left( p^{}_y \sigma^{}_x 
- p^{}_x \sigma^{}_y \right),
\label{H2}
\end{equation}
$\eta = +1$ ($K$ valley) and $-1$ ($K^{\prime }$ valley), $\tau^{}_{\alpha }$ and 
$\sigma^{}_{\alpha}$ are the Pauli matrices corresponding to the sublattices (A and B) and
the spin degrees of freedom, respectively. The parameters in Eqs.\ (\ref{H1})-(\ref{H2}) are: 
$v^{}_F$ is the Fermi velocity, $a$ is the lattice constant, $\lambda^{}_{\rm SO}$ is the SO 
coupling, and $\lambda^{}_R$ is the intrinsic Rashba SO coupling. For germanene and silicene 
these parameters are $a= 4.063$ \AA, $v^{}_F = 7.26\times 10^5$ m/s, $L^{}_z = 0.33$ \AA, 
$\lambda^{}_{\rm SO} = 43$ meV, $\lambda ^{}_R=10.7$ meV for germanene and $a= 3.866$ \AA, 
$v^{}_F = 8.47\times 10^5$ m/s, $L^{}_z = 0.23$ \AA, $\lambda^{}_{\rm SO} = 3.9$ meV, 
$\lambda^{}_R=0.7$ meV for silicene. The wave functions corresponding to the Hamiltonian 
(\ref{H1}) have four components of the form $\left(\psi^{}_{A\uparrow}, \psi^{}_{B\uparrow},
\psi^{}_{A\downarrow},\psi^{}_{B\downarrow} \right)$, where $\psi^{}_{A\alpha}$ 
and $\psi^{}_{B\alpha}$ determine the amplitude of the wave function in sublattice A and B, 
respectively, with spin direction $s = \uparrow, \downarrow$. In a magnetic field, the 
momentum $\vec{p}$ is replaced by the generalized momentum $\vec{\pi} = \vec{p} + e\vec{A}/c$, 
where $\vec{A}$ is the vector potential. To describe the LL wave functions, it is convenient to 
introduce the Landau functions of the nonrelativistic system $\phi^{}_{n\uparrow}$ ($\phi^{}_{n
\downarrow}$) in the corresponding nonrelativistic LL with index $n$ and spin direction $\uparrow$ 
or $\downarrow$. Then the structure of the LL wave functions in silicene/germanene 
can be schematically (without the coefficients) described as $(\phi^{}_{n\uparrow}, \phi^{}_{n+1
\uparrow}, \phi^{}_{n-1\downarrow},\phi^{}_{n\downarrow})$ ($K$ valley) and $(\phi^{}_{n+1\uparrow}, 
\phi^{}_{n\uparrow}, \phi^{}_{n\downarrow},\phi^{}_{n-1\downarrow})$ ($K^{\prime}$ valley). 

The FQHE is expected only in those LLs 
whose wave functions are mixtures of $\phi^{}_0$ and $\phi^{}_1$ \cite{FQHE_chapter,interaction}. There 
are two types of such LLs in silicene/germanene. The first type has the wave function of the form 
$(0, \phi^{}_{0\uparrow}, 0,0)$ (for K valley). This LL consists of only $\phi^{}_0$ and the interaction
in this LL is exactly the same as that in the $n=0$ nonrelativistic semiconductor system or
in graphene. The corresponding gaps are exactly the same as in a 
nonrelativistic system and they do not depend on the external electric field. 

\begin{figure}
\begin{center}\includegraphics[width=7cm]{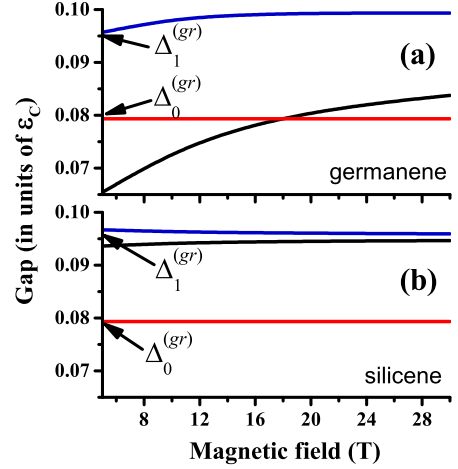}\end{center}
\vspace*{-1cm}
\caption{The $\nu = 1/3$ gap in three LLs corresponding to the Hamiltonian (\ref{HM}) as a 
function of the magnetic field. The electric field is zero. The results are shown for 
(a) germanene and (b) silicene. The color of the lines corresponds to the color of the LL shown 
in Fig.\ \ref{Fig_energy_E0_E01}(a). The 1/3 gaps of the graphene monolayer in the $n=0$ 
($\Delta _0^{\rm (gr)}$) and $n=1$ ($\Delta _0^{\rm (gr)}$) LLs are also shown. The finite size 
system has eight electrons.
}
\label{Fig_gap_E0}
\end{figure}

The second type of the wave functions has the form $\Psi^{}_2 = (C^{}_1 \phi^{}_{0\uparrow}, 
C^{}_2 \phi^{}_{1\uparrow}, 0,C^{}_3 \phi^{}_{0\downarrow})$ (we consider only the $K$ valley 
and include the coefficients $C^{}_1$, $C^{}_2$, and $C^{}_3$ in the wave functions). The wave 
functions $\Psi^{}_2$ effectively have three components and they are mixtures of the $n=0$ and 
$n=1$ nonrelativistic functions. In the basis of functions $\Psi^{}_2$ the Hamiltonian has 
the $3\times 3$ matrix form 
\begin{equation}
{\cal H}^{}_2 = \left(   
\begin{array}{ccc}
-\lambda^{}_{\rm SO} +S^{}_{+} &  \hbar \omega^{}_B & 0 \\
\hbar\omega^{}_B & \lambda^{}_{\rm SO} - S^{}_{-}  & - i \sqrt{2} (a/\ell^{}_0) \lambda^{}_R  \\
0 & i \sqrt{2} (a/\ell^{}_0) \lambda^{}_R  & -\lambda^{}_{\rm SO} - S^{}_{+}
\end{array}
\right)
\label{HM}
\end{equation}
where $\ell^{}_0 = (\hbar / eB )^{1/2}$ is the magnetic length, $\omega^{}_B = \sqrt{2} 
v^{}_F/\ell^{}_0$, and we have introduced the notations $S^{}_{+} = L^{}_zE^{}_z +\Delta^{}_z$, 
$S^{}_{-} = L^{}_zE^{}_z -\Delta^{}_z$. Here $\Delta^{}_z = g\mu^{}_B B$ is the Zeeman energy and 
we assumed that the $g$-factor in germanene/silicene is close to that of graphene $g\approx 2.2$.  
The eigenvalues and eigenvectors of the matrix Hamiltonian (\ref{HM}) determine the energies of three 
LLs and the corresponding wave functions (coefficients $C^{}_1$, $C^{}_2$, $C^{}_3$). 

\begin{figure}
\begin{center}\includegraphics[width=8cm]{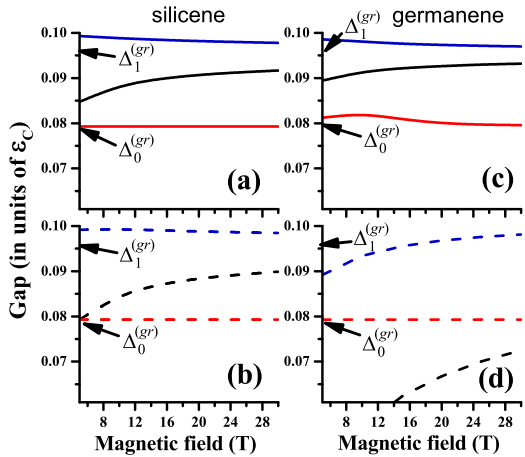}\end{center}
\vspace*{-1cm}
\caption{The $\nu = 1/3$ gap for three LLs corresponding to the Hamiltonian (\ref{HM}) as 
a function of the magnetic field. The electric field is $E^{}_z = 0.1$ V/\AA. The results 
are for (a,b) silicene and (c,d) germanene. The color of the lines corresponds to the color of the 
LL in Fig.\ \ref{Fig_energy_E0_E01}. The panels (a,c) correspond to the $K$ valley, while the
panels (b,d) correspond to the $K^{\prime }$ valley. The 1/3 gaps of the graphene monolayer
in the $n=0$ ($\Delta_0^{\rm (gr)}$) and $n=1$ ($\Delta_0^{\rm (gr)}$) LLs are also shown. 
}
\label{Fig_gap_E01}
\end{figure}

In the FQHE regime a given LL is partially occupied and the ground state of the electron system 
and the corresponding excitations are completely determined by the electron-electron interactions. 
The strength of the electron-electron interactions can be described by the Haldane's 
pseudopotentials, $V^{}_m$, \cite{haldane} which are the energies of two electrons with
relative angular momentum $m$. For the wave function $\Psi^{}_2 = (C^{}_1 \phi^{}_{0\uparrow}, 
C^{}_2 \phi^{}_{1\uparrow}, 0,C^{}_3 \phi^{}_{0\downarrow})$, which is characterized by the 
coefficients $C^{}_1$, $C^{}_2$, and $C^{}_3$, the Haldane's pseudopotentials are
\begin{equation}
V^{}_m = \int _0^{\infty } \frac{dq}{2\pi} q V(q)
\left[F (q) \right]^2 L^{}_m (q^2)
 e^{-q^2},
\label{Vm}
\end{equation}
where $L^{}_m(x)$ are the Laguerre polinomials, $V(q) = 2\pi e^2/(\kappa \ell^{}_0 q)$
is the Coulomb interaction in the momentum space, $\kappa$ is the
dielectric constant, and $F (q)$ is the corresponding form factor,
\begin{equation}
F(q) = \left( |C^{}_1|^2 + |C^{}_3|^2 \right)  L^{}_0\left( q^2/2 \right) +
|C^{}_2|^2 L^{}_{1}\left( q^2/2 \right) ,
\label{fn}
\end{equation}
To explore the correlation effects and the strength of electron-electron interactions in
a many electron germanene/silicene system we consider below the partially occupied LL with a
fractional filling factor corresponding to the FQHE \cite{spin_FQHE}. We study the many-electron 
system at fractional filling factors numerically within the spherical geometry \cite{haldane}. 
With the known Haldane's pseudopotentials (\ref{Vm}) we determine the interaction Hamiltonian 
matrix \cite{fano} and then numerically evaluate a few lowest eigenvalues and eigenvectors of
this matrix. The FQHE is observed when the ground state of the system is an incompressible liquid, 
the energy spectrum of which has a finite many-body gap. The magnitude of the gap indicates the 
interaction strength within a single LL and also determines the stability of the FQHE state. 
In conventional nonrelativistic system the FQHE is observed only in two lowest Landau levels, while 
in the higher Landau levels the charge density wave with gapless excitations has lower energy 
\cite{fogler97}. The most stable FQHE state, i.e., with the largest gap, is realized in
the $n=0$ LL in conventional nonrelativistic systems and in the $n=1$ LL in a graphene monolayer. 

In Fig.\ \ref{Fig_energy_E0_E01} the energy spectra, corresponding to the Hamiltonian (\ref{HM}) 
and consisting of three LLs, is shown for germanene as a function of the magnetic field. The results 
are for zero electric field [Fig.\ \ref{Fig_energy_E0_E01}(a)] and for $E^{}_z = 0.1$ 
V/\AA\, [Fig.\ \ref{Fig_energy_E0_E01} (b)]. For $E^{}_z = 0$, each LL has twofold valley degeneracy, 
which is lifted for a finite electric field. In Fig.\ \ref{Fig_energy_E0_E01}(b) the LL of
the $K$ and $K^{\prime }$ valleys are shown by solid and dashed lines, respectively. Without the SOI 
and for zero electric field, three LLs correspond to three LLs of graphene with energies 
$\varepsilon = 0$ ($n=0$ LL), and $\varepsilon =\pm \hbar \omega^{}_B \propto \pm 
\sqrt{B}$ ($n=\pm 1$ LLs). The SOI couples these states, which results in mixing of 
the corresponding wave functions and shifting of the energy levels, which is clearly seen in 
Fig.\ \ref{Fig_energy_E0_E01}. The level shown by a red line in Fig.\ \ref{Fig_energy_E0_E01}, and 
which originated from the $n=0$ LL of graphene, has a weak magnetic field dependence. For zero 
electric field the energy spectra can be found analytically and the three LLs shown in 
Fig.\ \ref{Fig_energy_E0_E01}(a) have energies $\varepsilon = \lambda^{}_{\rm SO}$ [red line in 
Fig.\ \ref{Fig_energy_E0_E01}(a)] and $\varepsilon = \pm \sqrt{ \lambda_{\rm SO}^2 + \hbar^2 
\omega_B^2 \left( 1+ a^2\lambda_R^2/\hbar^2 v_F^2 \right)}$ (black and blue lines). In this case 
the energy and the structure of the LL, shown by red line in Fig.\ \ref{Fig_energy_E0_E01}(a), do 
not depend on the magnetic field. The wave functions of this LL consist of  $\phi^{}_0$-type of 
nonrelativistic wave functions only. In a finite perpendicular electric field $E^{}_z$ [see 
Fig.\ \ref{Fig_energy_E0_E01}(b)], the LLs, shown by red lines in Fig.\ \ref{Fig_energy_E0_E01}(b), 
acquire a weak magnetic field dependence. Strong lifting of the valley degeneracy is also observed 
in Fig.\ \ref{Fig_energy_E0_E01}(b). The data in Fig.\ \ref{Fig_energy_E0_E01} are for germanene. 
For silicene, the results are similar but with a smaller energy scale due to the weaker SOI in 
silicene. It is convenient to label the LLs shown in Fig.\ \ref{Fig_energy_E0_E01} following the 
labeling scheme of graphene: $n=-1$, $n=0$, and $n=1$ LLs are shown by blue, red, and black lines, 
respectively. 

\begin{figure}
\begin{center}\includegraphics[width=8cm]{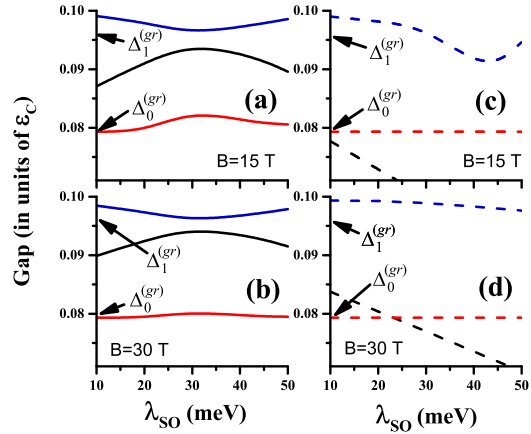}\end{center}
\vspace*{-1cm}
\caption{The $\nu = 1/3$ gap in three LLs corresponding to the Hamiltonian (\ref{HM}) as a 
function of the SOI ($\lambda^{}_{\rm SO}$). The electric field is $E^{}_z = 0.1$ V/\AA. The 
results are for germanene. The color of the lines corresponds to the color of the LL shown in 
Fig.\ \ref{Fig_energy_E0_E01}. The panels (a,b) correspond to $K$ valley, while the panels (c,d) 
correspond to $K^{\prime }$ valley. The magnetic field is (a,c) $B=15$ T and (b,d) $B=30$ T. The 
1/3-FQHE gaps of the graphene monolayer in the $n=0$ ($\Delta _0^{\rm (gr)}$) and the $n=1$ 
($\Delta _0^{\rm (gr)}$) LLs are also shown.
}
\label{Fig_gap_SO}
\end{figure}

The $\nu=1/3$ gaps for different LLs are shown in Fig.\ \ref{Fig_gap_E0} for zero electric 
field. The gaps $\Delta_0^{\rm (gr)}$ and $\Delta_1^{\rm (gr)}$ of graphene in the $n=0$ and 
the $n=1$ LLs are also shown. For the $n=0$ germanene/silicene LL, the gap is exactly the same 
as that of graphene ($\Delta_0^{\rm (gr)}$). The difference between the gaps in the $n=-1$ and 
$n=1$ LLs of germanene/silicene and the gap $\Delta_1^{\rm (gr)}$ of graphene illustrates the 
SO-induced coupling of the $n=-1$ and $n=1$ LLs. For silicene, the SO coupling is weak, which 
results in a small deviation of the gaps from the $\Delta_1^{\rm (gr)}$ value and a small 
splitting of the $n=-1$ and $n=1$ gaps. The strong SO coupling in germanene results in a large 
splitting of the gaps in the $n=-1$ and $n=1$ LLs [see Fig.\ \ref{Fig_gap_E0}(a). The difference
between the gaps in these LLs becomes smaller with increasing magnetic field. The FQHE 
in the $n=-1$ LL [blue line in Fig.\ \ref{Fig_gap_E0}(a)] is always greater than the largest 
gap $\Delta_1^{\rm (gr)}$ in graphene, while the gap in the $n=1$ LL [black line in 
Fig.\ \ref{Fig_gap_E0}(a)] is strongly suppressed, especially for small magnetic fields. Due to 
twofold valley degeneracy of the LLs in zero electric field, the behavior shown in Fig.\ 
\ref{Fig_gap_E0} is the same for both valleys, $K$ and $K^{\prime}$. 

In a finite electric field, the LLs in different valleys have different properties, which results 
in different values of the gaps. In Fig.\ \ref{Fig_gap_E01} the gaps in the LLs of 
germanene and silicene are shown as a function of the magnetic field for $E= 0.1$ V/\AA\, and 
different valleys. While the gap in the $n=0$ LL, similar to the case of $E^{}_z = 0$, has 
a weak magnetic field dependence, the gaps in the $n=1$ and $n=-1$ LLs are strongly modified 
compared to the $E^{}_z=0$ case. The changes in the behavior of the gaps are however quite 
different for silicene and germanene. 

For silicene [see Fig.\ \ref{Fig_gap_E01}(a,b)], the application of a perpendicular electric field 
strongly increases the difference between the values of the gaps in the $n=1$ and $n=-1$ LLs. 
The gap in the $n=-1$ LL is larger, while the gap in the $n=1$ LL is smaller than the 
gap $\Delta_1^{\rm (gr)}$ in graphene. The behavior of the gaps is the same for both valleys, 
while for the $K^{\prime}$ valley the gap in the $n=1$ LL is smaller than the one in the $K$ valley.
A different situation occurs for germanene, where the electric field 
reduces the difference between the gaps in the $n=-1$ and $n=1$ LLs for the $K$ valley and 
increase this difference for the $K^{\prime}$ valley [Fig.\ \ref{Fig_gap_E01}(c,d)]. The 
electric field also suppresses the gap in the $n=-1$ LL for the $K^{\prime }$ valley 
[Fig.\ \ref{Fig_gap_E01}(d)]. For small magnetic fields, $B\lesssim 15 $ T, the gap in
the $n=-1$ LL of the $K^{\prime}$ valley becomes even less than the gap $\Delta_1^{\rm (gr)}$. 
This behavior shows a strong sensitivity of the gaps on the magnitudes of the applied 
electric field and the magnetic field. 

The above results illustrate the importance of the SOI in determining the properties of the 
graphene-like systems. To illustrate the effect of the SOI on the value of the gaps, we vary 
the SO parameters, $\lambda^{}_{\rm SO}$, keeping all other parameters constant and equal to the 
parameters of germanene. The corresponding dependence of the gaps on $\lambda^{}_{\rm SO}$ is 
shown in Fig.\ \ref{Fig_gap_SO} for an electric field $E^{}_z = 0.1$ V/\AA\, and
for $B=15$ T and $B=30$ T. The gaps show clear nonmonotonic dependence on $\lambda^{}_{\rm SO}$ with 
a local maximum (minimum) at $\lambda^{}_{\rm SO}\approx 35$ meV. For the $K$ valley, both for 
the $n=-1$ and $n=1$ LLs the gaps are large and comparable to that in graphene. For $K^{\prime}$ 
the behavior is different. While for the $n=-1$ LL the gap is large and has a minimum at 
$\lambda^{}_{\rm SO}\approx 45$ meV (at $B=15$ T), the gap for the $n=1$ LL is strongly suppressed 
with increasing $\lambda^{}_{\rm SO}$. The suppression is stronger for smaller magnetic fields. 

In conclusion, we have shown that in graphene-like systems such as germanene and silicene, which 
have a strong SO interaction, there is an enhancement of the electron-electron interaction strength 
in one of the LL levels, which corresponds to the valence band of the system. This enhancement 
manifests itself as an increase of the gaps compared to that of graphene and is due to
the SO-induced coupling of the LLs of the conduction and valence bands, which modifies the 
corresponding wave functions and the interaction within a single LL. A perpendicular electric
field lifts the valley degeneracy of the systems and strongly modifies the interaction effects 
within a single LL. In one valley the electric field enhances the interaction strength 
(and the corresponding gaps) in the conduction band LL, while in another valley, the electric field 
strongly suppresses the interaction effects. 

The work has been supported by the Canada Research Chairs Program of the Government of Canada.

\end{document}